\documentclass[aps,pra,reprint,superscriptaddress]{revtex4-2}

\usepackage{siunitx} 		
\usepackage{bm} 			
\usepackage{amssymb}	 	
\usepackage{amsmath}		
\usepackage{hyperref}       
\usepackage{graphicx}		

\DeclareMathOperator{\gradop}{grad}
\DeclareMathOperator{\divop}{div}
\DeclareMathOperator{\trop}{tr}
\DeclareMathOperator{\Real}{Re}
\DeclareMathOperator{\Imag}{Im}

\begin{document}

\title{Thermoelastic Damping in MEMS Gyroscopes at High Frequencies}

\author{Daniel Schiwietz}
\email{daniel.schiwietz@de.bosch.com}
\affiliation{Robert Bosch GmbH, Corporate Research, 71272 Renningen, Germany}
\affiliation{Department of Electrical \& Computer Engineering, Technical University of Munich, 85748 Garching, Germany}

\author{Eva M. Weig}
\email{eva.weig@tum.de}
\affiliation{Department of Electrical \& Computer Engineering, Technical University of Munich, 85748 Garching, Germany}
\affiliation{Munich Center for Quantum Science and Technology (MCQST), 80799 Munich, Germany}
\affiliation{TUM Center for Quantum Engineering (ZQE), 85748 Garching, Germany}

\author{Peter Degenfeld-Schonburg}
\email{peter.degenfeld-schonburg@de.bosch.com}
\affiliation{Robert Bosch GmbH, Corporate Research, 71272 Renningen, Germany}

\date{\today}

\begin{abstract}
Microelectromechanical systems (MEMS) gyroscopes are widely used, e.g. in modern automotive and consumer applications, and require signal stability and accuracy in rather harsh environmental conditions. In many use cases, device reliability must be guaranteed under large external loads at high frequencies. The sensitivity of the sensor to such external loads depends strongly on the damping, or rather quality factor, of the high frequency mechanical modes of the structure. In this paper, we investigate the influence of thermoelastic damping on several high frequency modes by comparing finite element simulations with measurements of the quality factor in an application-relevant temperature range. We measure the quality factors over different temperatures in vacuum, to extract the relevant thermoelastic material parameters of the polycrystalline MEMS device. Our simulation results show a good agreement with the measured quantities, therefore proving the applicability of our method for predictive purposes in the MEMS design process. Overall, we are able to uniquely identify the thermoelastic effects and show their significance for the damping of the high frequency modes of an industrial MEMS gyroscope. Our approach is generic and therefore easily applicable to any mechanical structure with many possible applications in nano- and micromechanical systems.
\end{abstract}

\maketitle

\section{\label{sec:intro}Introduction}
Microelectromechanical systems (MEMS) gyroscopes are well established and indispensable in modern consumer and automotive electronics \cite{shaeffer_mems_2013,neul_micromachined_2007}. Especially in automotive applications, where gyroscopes operate in safety-critical systems, device reliability is of utmost importance. Functionality has to be ensured under various harsh environmental conditions \cite{acar_environmentally_2009} and the sensor signal stability has to be maintained despite many adverse linear and nonlinear effects \cite{saukoski_zero-rate_2007,nabholz_nonlinear_2019}. Most importantly, sensors have to withstand temperatures ranging from \SI{-40}{\degreeCelsius} to \SI{+120}{\degreeCelsius} and should be insensitive against external vibrations \cite{neul_micromachined_2007}. Therefore, the ability to predict the sensitivity of the sensor to such external conditions is crucial during MEMS design. In the past, vibrational robustness was mainly concerned with frequencies up to a few tens of kHz \cite{neul_micromachined_2007,liewald_100_2013}. State of the art applications, however, are faced with ever-increasing requirements. Among these requirements is the robustness against large external loads, at frequencies much higher than the operational frequency of the oscillatory gyroscope. High eigenfrequency modes, far beyond the operational frequency, can be decisive for the response of the sensor. The response of the corresponding high frequency modes is, among other quantities, determined by their quality factors. At typical pressures of around a few millibar, the quality factors of low frequency modes are known to be limited by gas damping \cite{chandorkar_limits_2008,frangi_near_2016}. However, to the authors' knowledge, there has been no exhaustive research on the damping of high frequency modes in MEMS gyroscopes. Known damping mechanisms that can contribute to the quality factors of MEMS resonators are gas damping, thermoelastic damping (TED), anchor losses, surface losses, material losses and Akhiezer damping \cite{lu_investigation_2021,candler_investigation_2003,imboden_dissipation_2014,ekinci_nanoelectromechanical_2005,frangi_near_2016,duwel_experimental_2003,rodriguez_direct_2018,yasumura_quality_2000,rodriguez_direct_2019}. The first three are usually considered as the dominant mechanisms in polysilicon MEMS resonators. Material losses are considered negligible for silicon, as it exhibits very linear material behavior, and surface losses are mainly relevant in nanoresonators \cite{ekinci_nanoelectromechanical_2005,candler_investigation_2003,imboden_dissipation_2014,yasumura_quality_2000}. Akhiezer damping is only expected to be relevant for frequencies above \SI{10}{\mega\hertz} \cite{rodriguez_direct_2019} and for very high quality factor and frequency ($Q$-$f$) products \cite{ghaffari_quantum_2013}.

In this paper, we compare measured and simulated quality factors of industrial MEMS gyroscopes over a wide range of eigenfrequencies. The aim of this work is to illuminate the significance of the thermoelastic damping contributions for high frequency modes. We show that thermolastic damping indeed limits the quality factor of high-frequency modes of the gyroscope and is thus crucial for the gyroscope's response to high frequency vibrations. 

In Section~\ref{sec:experimental} we introduce the MEMS devices and the measurement method. In Section~\ref{sec:numerics} the governing equations of thermoelasticity are introduced and an efficient method to simulate thermoelastic damping, based on the finite element method (FEM), is derived. We then verify the validity of our method in Section~\ref{sec:results}, by comparing our simulation results to measured data. Finally, in Section~\ref{sec:conclusion}, we summarize our results and conclude that thermoelastic damping is highly relevant for high frequency modes in our devices.

\section{\label{sec:experimental}Experimental Setup}
Two different industrial three-axis MEMS gyroscope designs (A and B), developed by Bosch, were investigated. The devices are made of polycrystalline silicon and are therefore assumed to exhibit isotropic material behavior. The designs were measured with two different scanning laser Doppler vibrometers (SLDV) from Polytec. The oscillation modes of the gyroscopes were excited in the linear regime by broadband signals (see details below). The measurements were performed on a dense grid of points over the structures (see details below) and the spectra of velocity and displacement were obtained from a fast Fourier transform at each point. The measured displacement maps obtained from the grid allowed to identify the vibrational modes by comparing with the simulated mode shapes. The quality factors of the modes were obtained from the linewidths of the resonance peaks. The frequency resolution of $\sim$\SI{1}{\hertz} was sufficient for an accurate resolution of the peaks.

Design A is an unencapsulated single chip, that was held at \SI{1}{\milli\bar} and \SI{25}{\degreeCelsius} inside a vacuum chamber. The excitation was realized dominantly in out-of-plane direction via a piezo-shaker. A chirp signal in the range from \SI{10}{\kilo\hertz} to \SI{2}{\mega\hertz} was applied to the piezo-shaker. The measurement was performed with a 1D SLDV. Therefore, only out-of-plane modes were detected for design A. The measurement was performed on a grid of around 400 points over the structure. The measured out-of-plane modes were semi-automatically matched to simulated modes. Although this is prone to errors, it enables the investigation of quality factor trends over a wide range of eigenfrequencies.

Design B was measured on the wafer and not encapsulated. The excitation of design B was realized electrostatically. The design contains a capacitive comb-drive as well as three different electrode pairs for capacitive sensing. Applying an electric broadband signal to one of the drive or sense electrode pairs, while grounding the remaining electrodes, enabled the excitation of various in-plane or out-of-plane modes. The applied signal was a pseudo-random broadband signal in the range from \SI{30}{\kilo\hertz} to \SI{1.25}{\mega\hertz}. The wafer containing design B was mounted on a thermal chuck inside a vacuum chamber. Thus, temperature and pressure could be varied. A 3D SLDV was used to measure in-plane and out-of-plane modes of design B. For the measurement of the out-of-plane modes a grid of around 100 points over the structure was used. The in-plane modes were measured inidividually with a grid of around 30 points each, which were only placed on the relevant oscillating parts of the structure. The measured modes of design B were manually identified with simulated modes, based on mode shapes and eigenfrequencies.

\begin{figure}[t]
	\includegraphics[keepaspectratio]{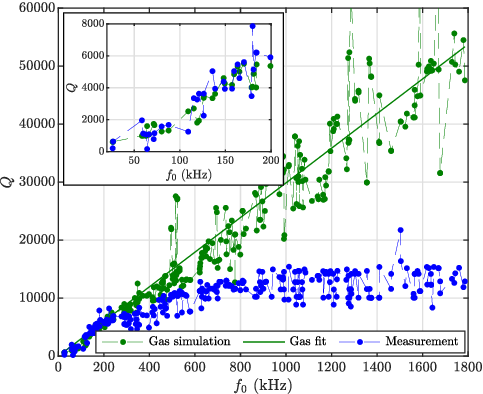}
	\caption{\label{fig:Qovermodesmotivation} Quality factors for all out-of-plane modes up to \SI{1.8}{\mega\hertz} of design A plotted over the eigenfrequencies of the modes at \SI{1}{\milli\bar} and \SI{25}{\degreeCelsius}. The green dots show the simulated quality factors based on gas damping. The solid green line is a linear fit of the simulated quality factors over frequency and indicates the trend of the gas damping quality factors. The blue dots show the measured quality factors. The inset shows a magnified plot up to \SI{200}{\kilo\hertz}.}
\end{figure}

Fig.~\ref{fig:Qovermodesmotivation} shows the measured quality factors of design A for several out-of-plane modes up to an eigenfrequency of \SI{1.8}{\mega\hertz} at \SI{1}{\milli\bar} and \SI{25}{\degreeCelsius}. The pressure of \SI{1}{\milli\bar} is a typical operational value for MEMS gyroscopes \cite{liewald_100_2013}. Additionally, simulated quality factors based on gas damping are also included in the figure. The gas damping simulations have been done using a Bosch internal gas damping simulation tool based on molecular flow simulations in COMSOL \cite{comsol_2017}. The validity and precision of the gas damping simulation is highlighted in the inset of Fig.~\ref{fig:Qovermodesmotivation} showing a closeup of the frequency regime up to 200 kHz. Up to \SI{200}{\kilo\hertz}, the measured quality factors follow the trend of the simulations. However, for higher eigenfrequencies the measured quality factors clearly saturate. Gas damping quality factors, on the other hand, increase approximately linearly with eigenfrequency \cite{ekinci_nanoelectromechanical_2005,chandorkar_limits_2008}. This motivates the incorporation of additional damping mechanisms into the simulation, to identify and accurately predict the damping contributions and the total quality factor. In this work, we will investigate the influence of TED on the two MEMS gyroscope designs.

\section{\label{sec:numerics}Numerical Analysis of Thermoelastic Damping}
Thermoelastic damping arises naturally from the coupling of the displacement and temperature fields. Therefore, any material with a non-zero thermal expansion coefficient exhibits TED. When a thermoelastic structure with a positive thermal expansion coefficient oscillates, regions under compression will heat up and regions under tension cool down. Thus, oscillatory temperature gradients arise across the structure. The resulting periodic heat flow along the temperature gradients is irreversible and leads to dissipation of energy. Zener pioneered the research on TED and derived an approximate analytic equation for the corresponding quality factor of a beam's fundamental bending mode \cite{zener_internal_1937,zener_internal_1938}. Lifshitz and Roukes later derived a refined solution for the same problem \cite{lifshitz_thermoelastic_2000}. In order to obtain quality factors for arbitrary geometries, the finite element method can be employed. It has been shown, that TED quality factors can be obtained from a complex eigenvalue problem of the thermoelastic equations \cite{antkowiak_design_2003} or from the calculation of dissipated and stored energy \cite{duwel_engineering_2006,hao_thermal-energy_2009,serra_finite_2009}.

We show how TED of industrial scale problems can be modelled by deriving a modified equation of motion for the mechanics, which can then be transferred into a mechanical reduced order model (ROM). In Section~\ref{sec:goveq} the governing differential equations of continuum mechanics are introduced along with their FEM formulation. In Section~\ref{sec:solution} an efficient simulation method for the evaluation of TED quality factors is derived.

\subsection{\label{sec:goveq}Governing Equations}
We start with the fundamental equations of thermoelasticity, which can be found e.g. in \cite{biot_thermoelasticity_1956,nowacki_thermoelasticity_1986}. The governing equation of the mechanical response, i.e. the equation of motion, is given by the linear momentum balance. For small deformations and in the absence of body forces it reads
\begin{equation}
	\divop(\bm{\sigma}) = \rho\bm{a}, \label{eq:eom}
\end{equation}
where $\bm{\sigma}$ is the stress tensor, $\rho$ is the density and $\bm{a}$ is the acceleration vector. The coupling to the temperature field affects Eq.~(\ref{eq:eom}) via thermal expansion. As this work is concerned with structures made of polysilicon, linear and isotropic material behavior will be assumed for mechanical and thermal properties. The constitutive equation accompanying Eq.~(\ref{eq:eom}) is given as
\begin{equation}
	\bm{\sigma} = \mathbb{C}[\bm{\varepsilon}-\bm{1}\alpha\Delta T], \label{eq:hooke}
\end{equation}
where $\mathbb{C}$ is the fourth-order elasticity tensor, $\bm{\varepsilon}$ is the total strain tensor, $\bm{1}$ is the second-order unit tensor, $\alpha$ is the thermal expansion coefficient and $\Delta T$ is the difference between the temperature field $T$ within the body and the ambient temperature $T_0$, i.e. $\Delta T = T - T_0$. The second term in the bracket of Eq.~(\ref{eq:hooke}) signifies the strain due to thermal expansion. The temperature changes that result from the thermoelastic coupling are generally very small. Therefore, the heat equation, which determines $\Delta T$, is linearized around $T_0$ as
\begin{equation}
	-\divop(\bm{q}_t) - T_0\alpha\trop(\bm{\dot{\sigma}}) = \rho C_V\Delta\dot{T}, \label{eq:heat}
\end{equation}
with heat flux vector $\bm{q}_t$, specific heat $C_V$ and time derivatives denoted by dots above the symbols. It is assumed that no additional heat sources are present within the body. The coupling to the stress field in Eq.~(\ref{eq:heat}) manifests itself in the heating of regions under compression and cooling of regions under tension, if the thermal expansion coefficient is positive. The constitutive equation for the heat flux vector is given by Fourier's law
\begin{gather}
	\bm{q}_t = -\kappa\gradop(\Delta T), \label{eq:fourier}
\end{gather}
where $\kappa$ is the thermal conductivity. 

The global FEM equations can be obtained in the usual way, by deriving and discretizing the weak forms of the local equations (\ref{eq:eom}) and (\ref{eq:heat}), leading to
\begin{gather}
	\bm{M}\bm{\ddot{u}} + \bm{K^u}\bm{u} + \bm{K^{ut}}\Delta\bm{T} = \bm{f}, \label{eq:femeom}
	\\
	\bm{C^t}\Delta\bm{\dot{T}} + \bm{K^t}\Delta\bm{T} = T_0(\bm{K^{ut}})^T\bm{\dot{u}}, \label{eq:femheat}
\end{gather}
where $\bm{M}$ is the mass matrix, $\bm{K^u}$ the stiffness matrix, $\bm{K^{ut}}$ the thermoelastic coupling matrix, $\bm{C^t}$ the specific heat matrix, $\bm{K^t}$ the thermal conductivity matrix, $\bm{u}$ and $\Delta\bm{T}$ are the nodal displacement and temperature change vectors and $\bm{f}$ is the external force vector. Only the oscillating structure is considered in the simulations. In Eqs.~(\ref{eq:femeom}) and (\ref{eq:femheat}) we assume that the displacement and temperature change are zero at the connection of the oscillating structure to the substrate. Furthermore, in Eq.~(\ref{eq:femheat}) we assumed insulating boundary conditions on the boundary that isn't fixed. See e.g. \cite{inc_ansys_2021} for the definitions of the FEM matrices.

\subsection{\label{sec:solution}Solution Method}
Several approaches exist to evaluate the thermoelastic damping of mechanical modes based on Eqs.~(\ref{eq:femeom}) and (\ref{eq:femheat}). Common but computationally expensive methods solve the coupled Eqs.~(\ref{eq:femeom}) and (\ref{eq:femheat}) simultaneously. However, these methods require the solution of non-symmetric equation systems with $4N$ degrees of freedom for a mesh with $N$ nodes. In the modelling of MEMS gyroscopes one usually deals with models where $N>10^6$ and quality factors have to be calculated for many modes over a wide frequency range. Therefore, solving the coupled problem of Eqs.~(\ref{eq:femeom}) and (\ref{eq:femheat}) is time consuming and numerically expensive. Instead, we will take a different approach, where we eliminate the heat equation and arrive at an effective equation of motion, which can then be efficiently evaluated in a ROM.

We consider the case where Eq.~(\ref{eq:femeom}) is harmonically driven at a frequency $\omega$. Thus, the steady-state oscillations of displacement and temperature change are given as
\begin{gather}
\bm{u}=\Real\{\bm{u}_0e^{i\omega t}\}, \label{eq:uansatz}
\\
\Delta\bm{T}=\Real\{\Delta\bm{T}_0e^{i\omega t}\}, \label{eq:tansatz}
\end{gather}
where $\bm{u}_0$ and $\Delta\bm{T}_0$ are the complex steady-state amplitudes. Equations~(\ref{eq:uansatz}) and (\ref{eq:tansatz}) are inserted into the heat equation (\ref{eq:femheat}), which can then be formally solved for $\Delta\bm{T}_0$. Consequently, one can then express the temperature change $\Delta\bm{T}$, based on Eq.~(\ref{eq:tansatz}), in dependence of displacement $\bm{u}$ and velocity $\bm{\dot{u}}$ as
\begin{equation}
	\Delta\bm{T}=-\omega T_0\Imag\{\bm{A}\}\bm{u}+T_0\Real\{\bm{A}\}\bm{\dot{u}}, \label{eq:tsol}
\end{equation}
where $\bm{A}=(\bm{K^t}+i\omega\bm{C^t})^{-1}(\bm{K^{ut}})^T$. Substituting Eq.~(\ref{eq:tsol}) into the equation of motion (\ref{eq:femeom}), we obtain the modified equation of motion
\begin{equation}
	\bm{M}\bm{\ddot{u}}+\bm{\tilde{C}}\bm{\dot{u}}+\bm{\tilde{K}}\bm{u}=\bm{f}, \label{eq:modeom}
\end{equation}
with damping matrix
\begin{equation}
	\bm{\tilde{C}}=T_0\Real\{\bm{K^{ut}}\bm{A}\}
\end{equation}
and stiffness matrix
\begin{equation}
	\bm{\tilde{K}}=\bm{K^u}-\omega T_0\Imag\{\bm{K^{ut}}\bm{A}\}.
\end{equation}
Note that Eq.~(\ref{eq:modeom}) is still exact in the sense that it fully incorporates the effect of the thermoelastic coupling on the mechanics for harmonic forcing. For oscillatory structures, such as MEMS gyroscopes, the equation of motion is usually solved in a modal ROM. The mechanical modes are obtained from the purely mechanical eigenvalue problem
\begin{equation}
(\bm{K^u}-\omega_n^2\bm{M})\bm{\phi}_n=\bm{0}, \label{eq:evp}
\end{equation}
with eigenfrequency $\omega_n$ and mode shape $\bm{\phi}_n$ of the $n$-th mode. The mode shapes are mass-normalized, i.e. $\bm{\phi}^T_n\bm{M}\bm{\phi}_n=1$.
The displacement is then expressed as a superposition of the modes
\begin{equation}
	\bm{u}\approx\bm{\Phi}\bm{q}, \label{eq:modsup}
\end{equation}
where $q_n$ is the modal coordinate of mode $n$ and $\bm{\Phi}=[\bm{\phi}_1 \ \bm{\phi}_2 \ ... \ \bm{\phi}_m]$ is a matrix, which contains the mass-normalized eigenvector of mode $n$ in the $n$-th column. The index $m$ indicates the mode at which the superposition is truncated, leading to an approximation of the actual $\bm{u}$. Inserting the modal superposition given by Eq.~(\ref{eq:modsup}) into Eq.~(\ref{eq:modeom}) and multiplying by $\bm{\Phi}^T$ from the left one obtains
\begin{equation}
	\bm{\ddot{q}}+\bm{\Phi}^T\bm{\tilde{C}}\bm{\Phi}\bm{\dot{q}}+\bm{\Phi}^T\bm{\tilde{K}}\bm{\Phi}\bm{q}=\bm{\Phi}^T\bm{f}. \label{eq:modeomrom}
\end{equation}
The effect of the thermoelastic coupling thus influences the mechanical modes by a damping contribution as well as a change in stiffness, i.e. a change of the eigenfrequencies. Furthermore, the modal damping and stiffness matrices $\bm{\Phi}^T\bm{\tilde{C}}\bm{\Phi}$ and $\bm{\Phi}^T\bm{\tilde{K}}\bm{\Phi}$ are not diagonal, i.e. they lead to a linear coupling between modes. This is simply a manifestation of the two-way coupling of Eqs.~(\ref{eq:femeom}) and (\ref{eq:femheat}). The temperature field that results from the motion of a mechanical mode and is determined by Eq.~(\ref{eq:femheat}) may also impose forces on other mechanical modes in Eq.~(\ref{eq:femeom}), providing an intermodal coupling. We assume that the effect of this coupling is weak and thus only consider the diagonal entries in Eq.~(\ref{eq:modeomrom}). Furthermore, the change of eigenfrequency due to thermoelastic coupling is very small and therefore only a very small error is made by neglecting it.

The damping matrix $\bm{\tilde{C}}$ depends on the oscillation frequency $\omega$. In this work, we are interested in the damping of a mode at its eigenfrequency $\omega_n$. Hence, to obtain the quality factor of mode $n$, one can set $\omega=\omega_n$. The reciprocal quality factor due to thermoelastic damping $Q_{TED,n}^{-1}$ is found by dividing the $n$-th diagonal entry of $\bm{\Phi}^T\bm{\tilde{C}}\bm{\Phi}$ by $\omega_n$, leading to
\begin{equation}
	\frac{1}{Q_{TED,n}}=\Real\left\{\frac{T_0}{\omega_n}\bm{\phi}_n^T\bm{K^{ut}}\left(\bm{K^t}+i\omega_n\bm{C^t})^{-1}(\bm{K^{ut}}\right)^T\bm{\phi}_n\right\}. \label{eq:invqted}
\end{equation}
Remarkably, Eq.~(\ref{eq:invqted}) allows us to determine the quality factors by only having to solve a symmetric linear equation system of size $N$, i.e. the size of the temperature degrees of freedom, per mode. Therefore, this approach is much more efficient than solving the coupled equations directly and is suitable for large models. We have implemented the assembly of the FEM matrices and the evaluation of Eq.~(\ref{eq:invqted}) in a self-written Matlab code.

We note that Eq.~(\ref{eq:invqted}) is equivalent to the result obtained with a perturbation method in \cite{bindel_structured_2006}. Furthermore, we note that the same expression can be obtained by calculating the quality factor as the ratio of stored to dissipated energy, if one calculates the dissipated energy due to the temperature field given by Eq.~(\ref{eq:tsol}) and neglects the effect of temperature on the stored energy.

\section{\label{sec:results}Results}
\begin{figure*}[ht!]
	\includegraphics[keepaspectratio]{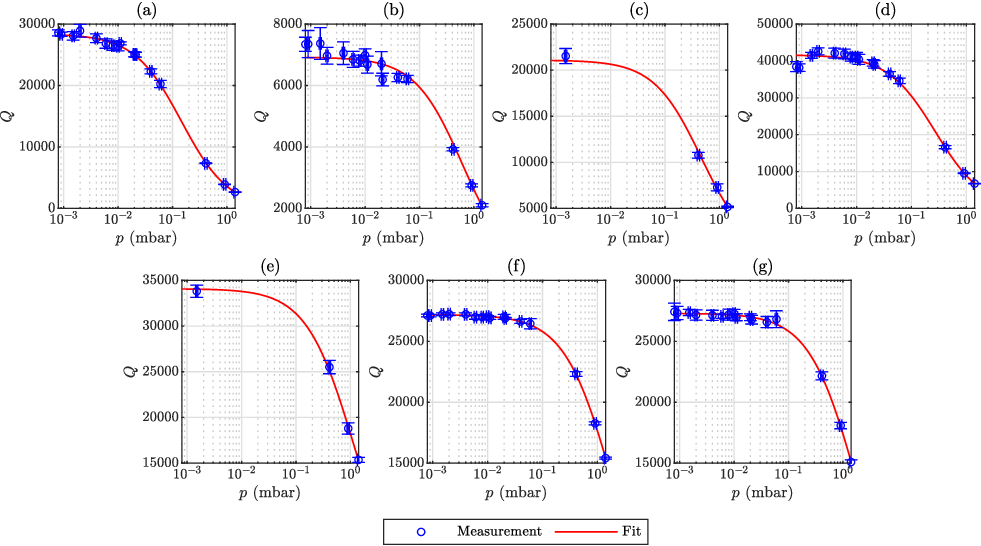}
	\caption{\label{fig:Qoverp} Quality factors over pressure for the 7 measured modes of design B, device 1. The measurements were performed at a temperature of \SI{20}{\degreeCelsius}. The blue circles show the experimental values. The red curves were obtained from a linear fit of the reciprocal quality factors, i.e. $Q^{-1}=m\cdot p+b$. The plots in (c) and (e) show in-plane modes, while the remaining measurements show out-of-plane modes. The corresponding modes and their simulated eigenfrequencies are: (a) Mode $a$ with $f_0=\SI{118.98}{\kilo\hertz}$, (b) Mode $b$ with $f_0=\SI{121.77}{\kilo\hertz}$, (c) Mode $c$ with $f_0=\SI{131.91}{\kilo\hertz}$, (d) Mode $d$ with $f_0=\SI{188.44}{\kilo\hertz}$, (e) Mode $e$ with $f_0=\SI{331.01}{\kilo\hertz}$, (f) Mode $f$ with $f_0=\SI{629.11}{\kilo\hertz}$, (g) Mode $g$ with $f_0=\SI{733.46}{\kilo\hertz}$.}
\end{figure*}
The main damping mechanisms in MEMS resonators are gas damping, thermoelastic damping and anchor losses. Other possible damping mechanisms include material losses and surface losses. Material losses are known to be negligible for silicon and surface losses are mainly relevant for nanoresonators \cite{candler_investigation_2003,imboden_dissipation_2014}. Additionally, Akhiezer damping has been observed in silicon MEMS resonators, but is only expected to be relevant for frequencies above \SI{10}{\mega\hertz} \cite{rodriguez_direct_2019} and for very high $Q$-$f$ products \cite{ghaffari_quantum_2013}.

From here on, when we refer to temperature, we mean the temperature $T_0$ of the atmosphere surrounding the oscillating part of the structure, i.e. the temperature inside the vacuum chamber.

Gas damping depends on temperature $T_0$ and pressure $p$, thermoelastic damping only depends on temperature and anchor losses are assumed to be independent of pressure and temperature. The total reciprocal quality factor is obtained from the sum of the reciprocal quality factors of the individual damping mechanisms
\begin{equation}
\frac{1}{Q(T_0,p)}=\frac{1}{Q_{\textrm{gas}}(T_0,p)} + \frac{1}{Q_\textrm{TED}(T_0)} + \frac{1}{Q_\textrm{anchor}}, \label{eq:invQ}
\end{equation}
with total quality factor $Q$, gas damping quality factor $Q_\textrm{gas}$, thermoelastic damping quality factor $Q_\textrm{TED}$ and anchor loss quality factor $Q_\textrm{anchor}$. In principal, as already mentioned, there are also other damping mechanisms that contribute to Eq.~(\ref{eq:invQ}). We assume that these other damping mechanisms are negligible compared to the gas damping, TED and anchor losses. We note, however, that other temperature- and pressure-independent damping mechanisms would not be distinguishable from anchor losses in our measurements. The dependence of $Q_\textrm{gas}$ on experimental conditions is particularly simple. At very low pressures, in the molecular regime, it scales as $Q^{-1}_\textrm{gas}\propto p$. At higher pressures, a transition into the viscous gas damping regime occurs, where the dissipation scales as $Q^{-1}_\textrm{gas}\propto \sqrt{p}$ \cite{ekinci_nanoelectromechanical_2005}. In the molecular regime, if the pressure isn't controlled, the dissipation scales with temperature as $Q^{-1}_\textrm{gas}\propto \sqrt{T_0}$ \cite{kim_temperature_2008}.

In order to verify that $Q_\textrm{TED}$ is determined by Eq.~(\ref{eq:invqted}), we measured 7 different modes of design B. In contrast to design A, which was excited via a piezo-shaker, design B was excited electrostatically. Due to the placement of the electrodes, only certain mode shapes were excitable. Thus, it wasn't possible to excite as many modes for design B as for the out-of-plane measurements of design A. Out of the measured modes, we chose those that could be identified unambiguously with simulated mode shapes and exhibited a clear resonance peak in our measurements. This lead to the 7 modes, which are enumerated by letters $a$ to $g$, from lowest to highest eigenfrequency. The lowest measured mode is mode $a$ with a simulated eigenfrequency of $f_0=\SI{118.98}{\kilo\hertz}$ and the highest measured mode is mode $g$ with a simulated eigenfrequency of $f_0=\SI{733.46}{\kilo\hertz}$. Out of the 7 measured modes, 2 are in-plane modes and the remaining 5 are out-of-plane modes.

\subsection{\label{sec:gasdamping}Gas Damping}
Since $Q^{-1}_\textrm{gas}\propto p$ in the molecular regime, gas damping can be made negligible by reducing the pressure sufficiently.
Figure~\ref{fig:Qoverp} shows the measured quality factors of the 7 measured modes of design B over pressure at a temperature of \SI{20}{\degreeCelsius}. For each pressure the quality factor of every mode was measured at 8 different spots on the MEMS structure. The spots were chosen individually for each mode according to the mode's anti-nodes. From the 8 measurements the mean value was calculated and the standard deviation was used for the vertical errorbars. Furthermore, a fit is shown, which was obtained for each mode from the linear relationship of the reciprocal quality factor and pressure, i.e $Q^{-1}=m\cdot p+b$, where $m$ is the slope and $b$ is the pressure-independent offset. It can be seen that all modes follow this expected trend, which confirms that the measurements were performed in the molecular regime. The quality factors only show very little pressure dependence below \SI{e-2}{\milli\bar}. Subsequent measurements were performed at \SI{e-3}{\milli\bar}, to ensure that the gas damping contribution is negligible and the measured quality factors are in good approximation equal to the contributions from thermoelastic damping and anchor losses. 

\begin{figure*}[ht!]
	\includegraphics[keepaspectratio]{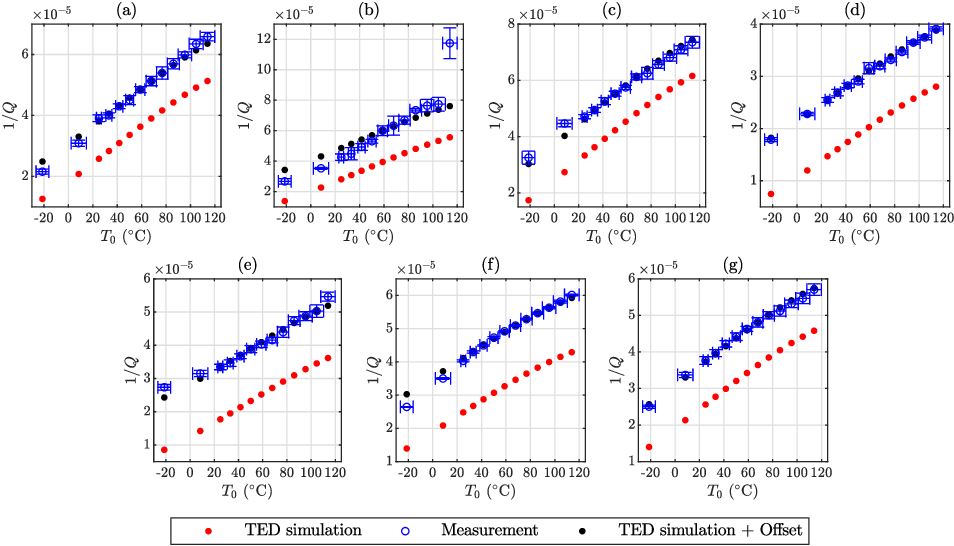}
	\caption{\label{fig:invQoverT} Reciprocal quality factors over temperature for the 7 measured modes of design B, device 2. The measurements were performed at a pressure of \SI{e-3}{\milli\bar}. The blue circles show the experimental values. The red dots were obtained from Eq.~(\ref{eq:invqted}) with temperature-dependent material properties as shown by the solid lines in Fig.~\ref{fig:materialparamsoverT}. The black dots were obtained by adding the mean difference between simulation and measurement to the simulated red dots, in order to emulate the effect of temperature-independent anchor losses. The plots in (c) and (e) show in-plane modes, while the remaining measurements show out-of-plane modes. The corresponding modes and their simulated eigenfrequencies are: (a) Mode $a$ with $f_0=\SI{118.98}{\kilo\hertz}$, (b) Mode $b$ with $f_0=\SI{121.77}{\kilo\hertz}$, (c) Mode $c$ with $f_0=\SI{131.91}{\kilo\hertz}$, (d) Mode $d$ with $f_0=\SI{188.44}{\kilo\hertz}$, (e) Mode $e$ with $f_0=\SI{331.01}{\kilo\hertz}$, (f) Mode $f$ with $f_0=\SI{629.11}{\kilo\hertz}$, (g) Mode $g$ with $f_0=\SI{733.46}{\kilo\hertz}$.}
\end{figure*}

\subsection{\label{sec:materialparameters}Material Parameters}
In order to verify $Q^{-1}_\textrm{TED}$ according to Eq.~(\ref{eq:invqted}), the correct temperature dependence has to be taken into account. At first sight Eq.~(\ref{eq:invqted}) appears to be linear in $T_0$. However, $Q^{-1}_\textrm{TED}$ also depends on thermal expansion coefficient $\alpha$, thermal conductivity $\kappa$ and specific heat $C_V$, which exhibit significant temperature dependencies. On the other hand, Young's modulus $E$, Poisson's ratio $\nu$ and density $\rho$ have much smaller temperature dependencies, which are negligible in this context. To make the dependence on temperature-dependent material properties more explicit, we rewrite Eq.~(\ref{eq:invqted}) as
\begin{widetext}
\begin{equation}
\frac{1}{Q_{TED,n}}=\Real\left\{\frac{\alpha^2 T_0}{\omega_n}\bm{\phi}_n^T\bm{\tilde{K}^{ut}}\left(\kappa\bm{\tilde{K}^t}+iC_V\omega_n\bm{\tilde{C}^t}\right)^{-1}\left(\bm{\tilde{K}^{ut}}\right)^T\bm{\phi}_n\right\}, \label{eq:invqted_2}
\end{equation}
\end{widetext}
where we defined $\bm{K^{ut}}=\alpha\bm{\tilde{K}^{ut}}$, $\bm{K^t}=\kappa\bm{\tilde{K}^t}$ and $\bm{C^t}=C_V\bm{\tilde{C}^t}$, so that $\bm{\tilde{K}^{ut}}$, $\bm{\tilde{K}^t}$ and $\bm{\tilde{C}^t}$ are then independent of $\alpha$, $\kappa$ and $C_V$. It is clear that Eq.~(\ref{eq:invqted_2}) scales with $\alpha^2$. Therefore, the thermal expansion coefficient $\alpha$ affects every mode in the same way. Thermal conductivity $\kappa$ and specific heat $C_V$, on the other hand, affect every mode in a different way, due to their appearance within the inverse matrix in Eq.~(\ref{eq:invqted_2}). To predict the quality factors accurately over temperature, the correct temperature dependencies of the material parameters have to be taken into account. For the purely mechanical properties, we assumed constant values of $E=\SI{161}{\giga\pascal}$, $\nu=0.22$ and $\rho=\SI{2330}{\kilo\gram\per\cubic\metre}$, which are the standard values used at Bosch for polycrystalline silicon. Due to a lack of reported data for polysilicon, the temperature-dependent specific heat $C_V$ was calculated from the Debye model with a Debye temperature of \SI{645}{\kelvin} for silicon \cite{kittel_introduction_2005}. The Debye model for silicon has also been used by others in the context of TED \cite{kim_temperature_2008}, albeit for monocrystalline silicon. We assume that the polycrystallinity has no significant impact on the specific heat. The value of $\kappa$ depends strongly on doping concentration and film thickness. Reported room temperature values for polysilicon samples of various doping concentrations and film thicknesses lie between \SI{15}{\watt\per\metre\per\kelvin} and \SI{60}{\watt\per\metre\per\kelvin} \cite{mcconnell_thermal_2005}. However, our samples have a film thickness of a few dozen micrometers, while the reported samples in \cite{mcconnell_thermal_2005} are significantly thinner. The thermal conductivity is expected to increase with film thickness and decrease with doping concentration \cite{mcconnell_thermal_2005}. Therefore, a thermal conductivity above \SI{60}{\watt\per\metre\per\kelvin} would be realistic for sufficiently low doping concentration. The thermal expansion coefficient $\alpha$ of monocrystalline silicon over temperature is well documented \cite{okada_precise_1984}. However, there exists no conclusive data for polycrystalline silicon. It has been suggested that the thermal expansion coefficient of polycrystalline silicon thin films might be significantly higher than that of bulk monocrystalline silicon \cite{tada_thermal_2000,tada_novel_2000}. Other researchers have performed measurements that found the thermal expansion coefficient of polycrystalline silicon to be constant over temperature and only slighty higher than that of monocrystalline silicon \cite{chae_measurement_1999}. Furthermore, it has been indicated in \cite{kahn_thermal_2002} that the thermal expansion coefficient of polycrystalline silicon differs from that of monocrystalline silicon depending on residual stresses. We conclude that there is ambiguous data on the temperature dependence of $\kappa$ and $\alpha$ for polysilicon thin films. Therefore, we will treat them as fit coefficients and estimate them from measured quality factors.

\begin{figure*}[ht!]
	\includegraphics[keepaspectratio]{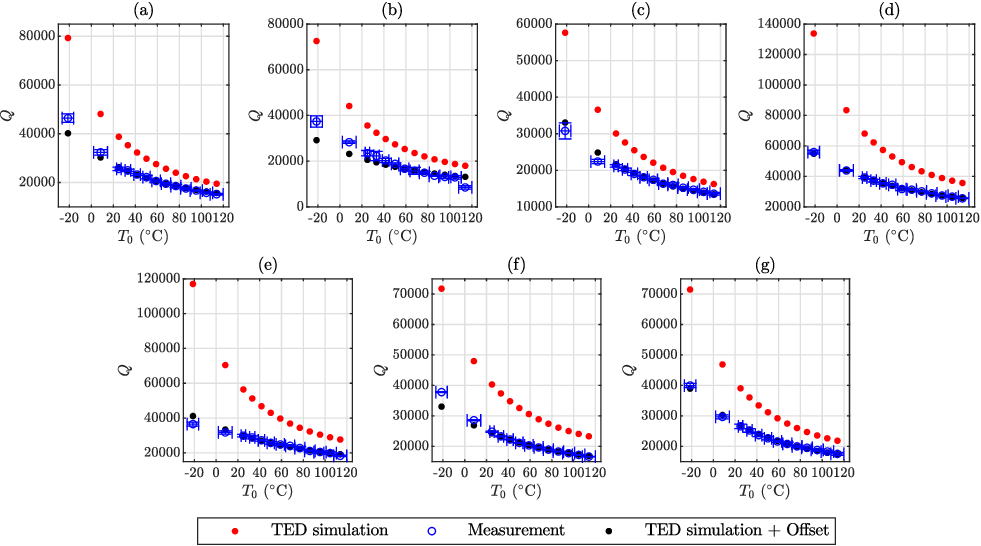}
	\caption{\label{fig:QoverT} Quality factors over temperature for the 7 measured modes of design B, device 2. The measurements were performed at a pressure of \SI{e-3}{\milli\bar}. The plots show the quality factors corresponding to the reciprocal values from Fig.~\ref{fig:invQoverT}. Blue circles show measured values, red dots show simulated TED and black dots show simulated TED with an additional temperature-independent damping contribution included. The plots in (c) and (e) show in-plane modes, while the remaining measurements show out-of-plane modes. The corresponding modes and their simulated eigenfrequencies are: (a) Mode $a$ with $f_0=\SI{118.98}{\kilo\hertz}$, (b) Mode $b$ with $f_0=\SI{121.77}{\kilo\hertz}$, (c) Mode $c$ with $f_0=\SI{131.91}{\kilo\hertz}$, (d) Mode $d$ with $f_0=\SI{188.44}{\kilo\hertz}$, (e) Mode $e$ with $f_0=\SI{331.01}{\kilo\hertz}$, (f) Mode $f$ with $f_0=\SI{629.11}{\kilo\hertz}$, (g) Mode $g$ with $f_0=\SI{733.46}{\kilo\hertz}$.}
\end{figure*}

Quality factors of design B were measured over temperature at a pressure of \SI{e-3}{\milli\bar}. The measurements span the range from $\SI{-21}{\degreeCelsius}$ to $\SI{114}{\degreeCelsius}$. The bond pads of the device that was measured over pressure were already damaged from contacting them multiple times. Therefore, we performed the temperature measurements on a second device of design B on the same wafer.
Due to the low pressure, the measured $Q^{-1}$ is approximately equal to the contributions from TED and anchor losses. We added the mean difference between simulated $Q^{-1}_\textrm{TED}$ and measured $Q^{-1}$ for each mode to $Q^{-1}_\textrm{TED}$, to emulate the effect of temperature-independent anchor loss contributions $Q^{-1}_\textrm{anchor}$. We then determined $\alpha$ and $\kappa$ such that the simulated $Q^{-1}_\textrm{TED}$ over temperature matched the measured $Q^{-1}$ after the mean difference was added. The measured $Q^{-1}$ (blue circles), simulated $Q^{-1}_\textrm{TED}$ (red dots) according to Eq.~(\ref{eq:invqted}) and $Q^{-1}_\textrm{TED}$ with added offset (black dots) are shown in Fig.~\ref{fig:invQoverT} over temperature. It can be seen that the simulated $Q^{-1}_\textrm{TED}$ is lower than the measured $Q^{-1}$ for all modes, as expected. The corresponding $Q$ values are shown in Fig.~\ref{fig:QoverT}, revealing that the $Q$ changes by a factor of $\sim$2 over the application-relevant temperature range.\\
In order to find suitable values for $\alpha$ and $\kappa$, we assumed simple temperature-dependencies. In our temperature range, the thermal expansion coefficient $\alpha$ of monocrystalline silicon increases with a declining slope over temperature \cite{okada_precise_1984}. We assume a qualitatively similar behavior for polycrystalline silicon. For simplicity, we choose a quadratic function for $\alpha(T_0)$
\begin{equation}
\alpha(T_0) = \alpha_0 + \alpha_1(T_0 - T_{RT}) - \alpha_2(T_0 - T_{RT})^2, \label{eq:alphaansatz}
\end{equation}
where $\alpha_0$, $\alpha_1$ and $\alpha_2$ are positive fit parameters and $T_{RT}=\SI{25}{\degreeCelsius}$. The thermal conductivity $\kappa$ of silicon tends to decrease with $1/T_0$ in our temperature range \cite{mcconnell_thermal_2005}. To mimic this behavior, we choose the fit function for $\kappa(T_0)$ in analogy to \cite{prakash_thermal_1978} as
\begin{equation}
	\kappa(T_0)=\kappa_0\left[1-\kappa_1\left(\frac{T_0 - T_{low}}{T_0}\right)\right], \label{eq:kappaansatz}
\end{equation}
where $\kappa_0$ and $\kappa_1$ are positive fit parameters and $T_{low}=\SI{-21}{\degreeCelsius}$ is the lowest measured temperature. The material parameters were estimated from least-square fits. For that purpose, mode $b$ was excluded, as it showed irregular behavior. The measured quality factor of mode $b$ was three times larger for the device that was measured over temperature than for the device that was measured over pressure, as can be seen from Fig.~\ref{fig:QoverT}~(b) and Fig.~\ref{fig:Qoverp}~(b). The devices are nominally identical and one would expect similar quality factors, as was the case for all other modes. Furthermore, the quality factor measurement of mode $b$ showed a clear outlier over temperature, as seen in Fig.~\ref{fig:invQoverT}~(b) and Fig.~\ref{fig:QoverT}~(b) at \SI{114}{\degreeCelsius}. We performed a measurement of mode $b$ at {\SI{e-3}{\milli\bar}} and {\SI{25}{\degreeCelsius}} on a third device and obtained a quality factor of $27000\pm1000$. This is closer to the value of $23000\pm1200$ in Fig.~\ref{fig:QoverT}~(b). Although we can not explain this peculiarity, we expect it to not be related to thermoelastic damping.

\begin{figure}[t]
	\includegraphics[keepaspectratio]{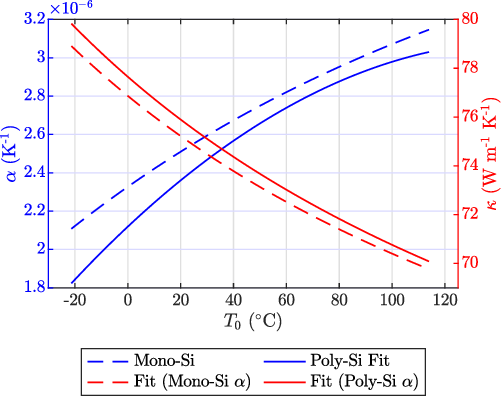}
	\caption{\label{fig:materialparamsoverT} Thermal expansion coefficient $\alpha$ and thermal conductivity $\kappa$ over  temperature. The dashed blue line shows the $\alpha$ of monocrystalline silicon according to \cite{okada_precise_1984} and was used as an initial guess. Subsequent material parameters were obtained by least-square fitting the black dots to the measured blue circles in Fig.~\ref{fig:invQoverT}. The dashed red line shows an initial fit for $\kappa$. The red and blue solid lines show the converged result of our fitting method for the thermal conductivity $\kappa$ and the thermal expansion coefficient $\alpha$, respectively.}
\end{figure}

In order to estimate the five unknown parameters $\alpha_0$, $\alpha_1$, $\alpha_2$, $\kappa_0$ and $\kappa_1$, we proceeded as follows: As a first estimate, we assumed the $\alpha(T_0)$ of monocrystalline silicon according to \cite{okada_precise_1984}, which is shown as the blue dashed line in Fig.~\ref{fig:materialparamsoverT}.
We then chose $\kappa(T_0)$ according to Eq.~(\ref{eq:kappaansatz}) and determined $\kappa_0$ and $\kappa_1$ from a least-square fit. For that purpose, we added the mean difference between measurement and simulation to the simulated $Q^{-1}_\textrm{TED}$ of each mode, to account for anchor losses. We then calculated the squared differences between the resulting values and the measured $Q^{-1}$ values. The squared differences for all modes over all temperatures, except mode $b$, were then added and the material parameters were determined from the minimization of this sum. The $\kappa(T_0)$ that was obtained from this first fit, based on the $\alpha(T_0)$ of monocrystalline silicon, is shown as a red dashed line in Fig.~\ref{fig:materialparamsoverT}. Subsequently, this $\kappa(T_0)$ was assumed and $\alpha(T_0)$ was chosen as in Eq.~(\ref{eq:alphaansatz}). The coefficients $\alpha_0$, $\alpha_1$ and $\alpha_2$ were then determined from a least-square fit, similar to the previous one, only differing in the unknown parameters. The resulting $\alpha(T_0)$ is shown as a solid blue line in Fig.~\ref{fig:materialparamsoverT} and can be seen to be lower than the $\alpha(T_0)$ of monocrystalline silicon. Finally, to obtain a corrected estimate for $\kappa(T_0)$, we assumed this new $\alpha(T_0)$ and treated $\kappa_0$ and $\kappa_1$ as fit parameters again. From this final fit we then obtained the $\kappa(T_0)$ according to the solid red line in Fig.~\ref{fig:materialparamsoverT}. For verification, we performed a final iteration, where we assumed $\kappa(T_0)$ according to the solid red line in Fig.~\ref{fig:materialparamsoverT} and again obtained the $\alpha(T_0)$ according to the solid blue line. Therefore, we conclude that the fitting process has converged to some minimum. The final material parameters, according to the solid red and blue lines in Fig.~\ref{fig:materialparamsoverT}, will be assumed for the remainder of this paper. Figures~\ref{fig:invQoverT} and \ref{fig:QoverT} show the $Q_\textrm{TED}$ based on these final material parameters.

\subsection{\label{sec:dampingcontributions}Damping Contributions}
If $\alpha(T_0)$ were to be rather constant over temperature, as suggested in \cite{chae_measurement_1999}, then the simulated values would not be steep enough to match the measured $Q^{-1}$ slope over $T_0$. The $\alpha(T_0)$, that was determined via fitting, leads to a good agreement with the measured $Q^{-1}$ over $T_0$ slope and is in a reasonable range compared to the monocrystalline values. The $\kappa(T_0)$ values are higher than the reported values in \cite{mcconnell_thermal_2005}, but seem reasonable due to the few dozen micrometer thickness of our samples. The change of around \SI{10}{\watt\per\metre\per\kelvin} of $\kappa(T_0)$ over our temperature range is similar to the data in \cite{mcconnell_thermal_2005}. However, we note that there might be other possible combinations of $\alpha_0$, $\alpha_1$, $\alpha_2$, $\kappa_0$ and $\kappa_1$, which also lead to good agreement with the measured data. In prinicipal, this is related to the unknown temperature-independent anchor loss contributions, which manifest themselves as an offset between $Q^{-1}_\textrm{TED}$ and $Q^{-1}$. There might be other curves for $\alpha(T_0)$ and $\kappa(T_0)$ that yield a similar slope of $Q^{-1}_\textrm{TED}$ over $T_0$ and only differ in their offset to $Q^{-1}$. Additionally simulating the anchor losses might help to circumvent this issue and will be subject of a future publication. Nevertheless, we conclude that the temperature dependence of the measured $Q^{-1}$ can be very well reproduced by the $Q^{-1}_\textrm{TED}$ according to Eq.~(\ref{eq:invqted}) with reasonable material parameters. This clearly shows the validity of Eq.~(\ref{eq:invqted}) and confirms the assumption that the temperature dependence of Eq.~(\ref{eq:invQ}), at pressures where gas damping is negligible, is due to $Q^{-1}_\textrm{TED}$ for the investigated MEMS gyroscope. 
In Fig.~\ref{fig:QoverT} it can be seen that the $Q$ at \SI{e-3}{\milli\bar} increases by roughly a factor of two from the highest to the lowest temperature. The temperatures are all within the application-relevant temperature range. To ensure device functionality over the whole temperature range, knowledge of quality factors is crucial. The significant change over temperature highlights the relevance of TED for the development of MEMS gyroscopes.
\begin{figure}
	\includegraphics[keepaspectratio]{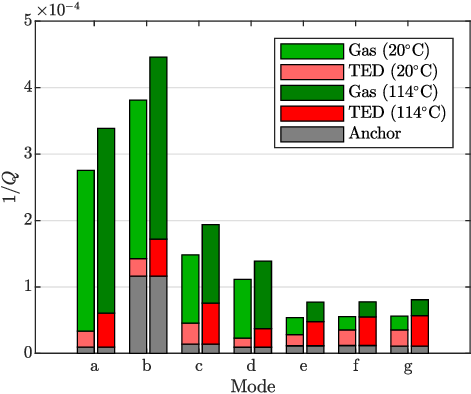}
	\caption{\label{fig:invQcomparison} Contributions of gas damping, TED and anchor losses to the total $Q^{-1}$ for the 7 modes of design B. The $Q^{-1}_\textrm{gas}$ was obtained from the fit in Fig.~\ref{fig:Qoverp} at \SI{1}{\milli\bar} and \SI{20}{\degreeCelsius} (light green). Assuming non-constant pressure it was then scaled to \SI{114}{\degreeCelsius} (dark green), according to $Q_\textrm{gas}^{-1}\propto\sqrt{T_0}$. $Q^{-1}_\textrm{TED}$ was obtained from Eq.~(\ref{eq:invqted}) for \SI{20}{\degreeCelsius} (light red) and \SI{114}{\degreeCelsius} (dark red). The temperature- and pressure-independent $Q^{-1}_\textrm{anchor}$ (gray) was calculated by subtracting $Q^{-1}_\textrm{TED}$ at \SI{20}{\degreeCelsius} from the y-intercept of the linear fit for $Q^{-1}$, that was shown in Fig.~\ref{fig:Qoverp}.}
\end{figure}

In Fig.~\ref{fig:invQcomparison} the $Q^{-1}$ contributions are shown for a pressure regime where gas damping is relevant. For that purpose, the contributions were obtained from the fit $Q^{-1}=m\cdot p+b$ in Fig.~\ref{fig:Qoverp}. The fit was obtained at $\SI{20}{\degreeCelsius}$. Therefore, one can calculate $Q_\textrm{gas}^{-1}(\SI{20}{\degreeCelsius},p)=m\cdot p$ and $Q^{-1}_\textrm{anchor}=b-Q^{-1}_\textrm{TED}(\SI{20}{\degreeCelsius})$. Since $Q_\textrm{gas}^{-1}\propto\sqrt{T_0}$, if the pressure isn't held constant, the gas damping can be calculated for other temperatures based on the known value at $\SI{20}{\degreeCelsius}$. This is relevant if the device would be encapsulated, where the pressure would also change over temperature. Based on Fig.~\ref{fig:Qoverp}, we extracted the gas damping at \SI{1}{\milli\bar} and \SI{20}{\degreeCelsius} and then scaled it up to \SI{114}{\degreeCelsius}, which also corresponds to a higher pressure. $Q^{-1}_\textrm{TED}$ is also shown for \SI{20}{\degreeCelsius} and \SI{114}{\degreeCelsius}, based on Eq.~(\ref{eq:invqted}). $Q^{-1}_\textrm{anchor}$ is temperature-independent and therefore identical for both temperatures. Looking at Fig.~\ref{fig:invQcomparison}, one can see that modes $a$ to $d$ are gas damping dominated. However, TED gains significance at the higher temperature of $\SI{114}{\degreeCelsius}$. This is due to the fact that, at least in our temperature range and with our devices, $Q^{-1}_\textrm{TED}$ scales roughly linear with $T_0$, as seen in Fig.~\ref{fig:invQoverT}, while $Q_\textrm{gas}^{-1}$ only scales with $\sqrt{T_0}$. On the other hand, for the higher modes $e$ to $g$, one can see that even at $\SI{20}{\degreeCelsius}$ the $Q^{-1}_\textrm{TED}$ is comparable to $Q_\textrm{gas}^{-1}$ and for modes $f$ and $g$ even larger than $Q_\textrm{gas}^{-1}$. At $\SI{114}{\degreeCelsius}$ this becomes even more pronounced. The $Q_\textrm{anchor}$ corresponding to Fig.~\ref{fig:invQcomparison} are between 70000 and 110000. The only exception is mode $b$ with a $Q_\textrm{anchor}$ of around 8500. Again, we note that the $Q$-factor of mode $b$ from the temperature and pressure measurements, which were performed on different but nominally identical devices, differ significantly. One can also obtain $Q^{-1}_\textrm{anchor}$ as the offset in Fig.~\ref{fig:invQoverT}. This leads to comparable values as the ones in Fig.~\ref{fig:invQcomparison}, except for mode $b$, for which the extracted $Q_\textrm{anchor}$ would then be 48000. Mode $b$ is a symmetric mode, whereas mode $a$ is the corresponding anti-symmetric mode. Therefore, the deviation for mode $b$ can not be attributed to a simple measurement error, as the entire spectrum was measured simultaneously and one would then expect a discrepancy for at least mode $a$ as well. We also exclude that a MEMS device was faulty, as we would then also expect to see significant deviations for the other modes. One possible explanation could be a substrate-related effect, which was only present in one of the two devices.
\begin{figure*}[ht!]
	\includegraphics[keepaspectratio]{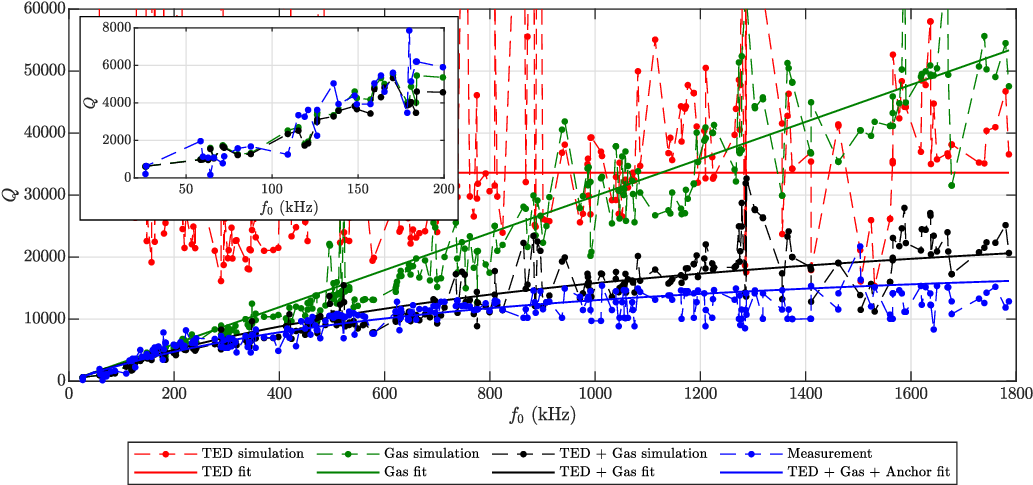}
	\caption{\label{fig:Qovermodes} Quality factors for several out-of-plane modes of design A plotted over the eigenfrequencies of the modes at \SI{1}{\milli\bar} and \SI{25}{\degreeCelsius}. The green dots show the simulated quality factors based on gas damping. The red dots show the simulated quality factors based on TED, i.e. Eq.~(\ref{eq:invqted}). The black dots show the resulting quality factors based on the gas damping and TED simulations. The blue dots show the measured quality factors. The solid lines indicate the trends of the data. The solid green line is a linear fit of the simulated gas damping quality factors over eigenfrequency. The solid red line was obtained as the median of the TED quality factors. The solid black line is the resulting quality factor due to the solid green and red lines. The solid blue line is equal to the solid black line with an additional constant $Q_\textrm{anchor}=75000$ included. The inset shows a magnified plot until \SI{200}{\kilo\hertz} without trend lines.}
\end{figure*}
\subsection{\label{sec:broadspectrum}Damping Contributions over a Wide Frequency Spectrum}
The same quality factors of design A as in Fig.~\ref{fig:Qovermodesmotivation} are shown again in Fig.~\ref{fig:Qovermodes}. The quality factors of several out-of-plane modes, measured at \SI{1}{\milli\bar} and \SI{25}{\degreeCelsius}, are shown as blue dots. The simulated gas damping quality factors are shown as green dots. Additionally, the simulated TED quality factors are included as red dots. Design A is made of the same material and features the same out-of-plane thickness as design B. Therefore, the material parameters according to the solid blue and red lines in Fig.~\ref{fig:materialparamsoverT} at \SI{25}{\degreeCelsius} were used. The quality factors resulting from gas damping plus TED are shown as black dots. Furthermore, trend lines are included as a guide to the eye. The red horizontal line signifies that there is no clear up- or downward trend for TED quality factors over the measured range. It was calculated as the median of the simulated TED quality factors, to avoid the bias by large outliers. The green line is a linear fit of the simulated gas damping quality factors over frequency. The black line was obtained from the previous two trend lines, i.e. based on the trend of TED and gas damping. Additionally, in order to estimate possible anchor losses, a constant $Q_\textrm{anchor}$ of 75000 was added to the contributions of the black line, to obtain the blue line. 

One can see that up to \SI{200}{\kilo\hertz} the influence of TED is rather small compared to gas damping, as $Q_\textrm{TED}$ is much larger than $Q_\textrm{gas}$. Therefore, $Q$ is well approximated by $Q_\textrm{gas}$ alone. However, the incorporation of TED into the model significantly decreases the simulated $Q$ for higher frequencies. For many modes this leads to a reduction of the simulated $Q$ by a factor of around 2. In fact, the simulation based on gas damping and TED is quite close to the measurements up to around \SI{800}{\kilo\hertz}. Above \SI{800}{\kilo\hertz} almost all measured modes exhibit a lower quality factor in the measurement than in the simulation. For the highest modes the measurement is still around 25\% below the simulation of gas damping and TED. For the modes above \SI{800}{\kilo\hertz} anchor losses play an increasing role. TED quality factors show no clear up- or downward trend over our measurement range. Gas damping quality factors increase approximately linearly with frequency. Therefore, for sufficiently high modes the gas damping quality factors approach the value of the TED quality factors and the effect of TED becomes relevant. For the modes above \SI{800}{\kilo\hertz}, the gas damping quality factors are large enough, such that further damping mechanisms, which tend to have larger quality factors than TED, become relevant as well. For illustration, a $Q_\textrm{anchor}$ of 75000 was therefore included in the blue trend line. The blue trend line then leads to a stronger saturation of $Q$ over mode frequency compared to only including gas damping and TED and therefore gives a better approximation of the measured data at high frequencies. However, this is only a phenomelogical remedy and it could also be that $Q_\textrm{anchor}$ has some trend over frequency. Clarifying the influence of anchor losses and investigating whether the incorporation of an anchor loss simulation helps to explain the remaining deviation between simulation and measurement at high frequencies will be part of future research. 

Note that the simulated $Q_\textrm{gas}$ is a simplified approximation. Furthermore, the measured data of design A in Fig.~\ref{fig:Qovermodes} is not as accurate as the measurements on design B. This is due to the fact that measuring and identifying as many modes as in Fig.~\ref{fig:Qovermodes} is very elaborate and the semi-automated evaluation of the quality factors of design A from the measured data is prone to errors. Therefore, deviations between measurement and simulation are to be expected for individual modes. Regardless, the measurement in Fig.~\ref{fig:Qovermodes} shows a clear trend. The simulations and the measured trend over the broad spectrum clearly show the significance of TED with increasing frequency. For higher temperatures or lower pressures TED would become even more important. The relevance of TED for higher modes is also in agreement with the observations made from design B, as was shown in Fig.~\ref{fig:invQcomparison}.

\section{\label{sec:conclusion}Conclusion}
We reported quality factor measurements of an industrial MEMS gyroscope (design A) for a multitude of out-of-plane modes over a wide frequency range. Although gas damping matches the observations for the first few modes, up to \SI{200}{\kilo\hertz}, we found that for higher modes the measured quality factors are significantly lower than the pure gas damping model predicts. The measured quality factors saturate for high frequency modes above \SI{800}{\kilo\hertz}, while gas damping quality factors increase approximately linearly with eigenfrequency. The deviation starts to become notable above \SI{200}{\kilo\hertz}. In order to account for this deviation, we introduced thermoelastic damping into our model. 

We demonstrated an efficient way to simulate thermoelastic damping, by eliminating the heat equation and deriving an effective equation of motion for the harmonically driven mechanics. The FEM equations were implemented in a self-written code. In order to verify our thermoelastic damping simulations, we measured the quality factors for 7 different modes of a second MEMS gyroscope design (design B) over temperature in a vacuum chamber. We found a good agreement between simulated and measured quality factors over temperature. We showed that the temperature dependence of thermoelastic material parameters, in particular the thermal expansion coefficient, has to be taken into account to reproduce the observed behavior over temperature. Fitting our simulation results to the measurements, we were able to estimate the thermal conductivity and thermal expansion coefficient over temperature. However, we note that the presence of damping mechanisms which are independent of pressure and temperature, such as anchor losses, leads to some uncertainty of the fitted material parameters.

Having validated the thermoelastic damping simulations and having obtained the relevant material parameters, we then applied our method to the measurement over a wide frequency range of gyroscope design A. Taking thermoelastic damping and gas damping into account, we then found good agreement with measured quality factors up to \SI{800}{\kilo\hertz}. For even higher frequencies, we found that additional damping mechanisms might be relevant, as the simulated quality factors were still higher than the measured ones. Nevertheless, the simulated quality factors were significantly closer to the measured ones over the entire frequency range, as compared to pure gas damping simulations.

Our results clearly show the significance of thermoelastic damping for high frequency modes in MEMS gyroscopes. On the lower end of the mode spectrum, gas damping dominates and is the sole relevant damping mechanism. However, for increasing eigenfrequency, thermoelastic damping gains relevance and appears to be the limiting damping mechanism, i.e. it has the smallest quality factor. At the upper end of the measured spectrum, additional damping mechanisms might have to be taken into account, although thermoelastic damping still seems to be the primary contribution. Expanding our model with anchor loss simulations, in order to improve the accuracy and also investigate the origin of the additonal damping contributions at high eigenfrequencies, will be subject of future research. Finally, as our method is based on FEM, it can be readily applied to other MEMS and NEMS structures where quality factors and damping mechanisms are also subject of ongoing research \cite{nabholz_amplitude-_2018,bhugra_quality_2017,tsaturyan_ultracoherent_2017}. We would like to emphasize that, although our discussion was focused on high frequency modes, our method is generally applicable also for low frequency modes.

\begin{acknowledgments}
The authors are thankful to Christian Budak at Robert Bosch GmbH for providing the measurements of design A. Furthermore, the authors gratefully acknowledge technical support from Ulrich Kunz and Thomas Buck at Robert Bosch GmbH for the measurements of design B.
\end{acknowledgments}

\bibliography{TED_refs}
\end{document}